# Surface-state mediated spin-to-charge conversion in Sb films via bilateral spin current injection

E. Gomes,[1] J. E. Abrão,[1] E. S. Santos,[1] S. Bedanta,[2] H. Ding,[3] J. B. S. Mendes,[4] and A. Azevedo[1,a)]

[1] Departamento de Física, Universidade Federal de Pernambuco, Recife, Pernambuco 50670-901, Brasil

[2] Laboratory for Nanomagnetism and Magnetic Materials (LNMM), School of Physical Sciences, National Institute of Science Education and Research (NISER), An OCC of Homi Bhabha National Institute (HBNI), Jatni-752050, India

[3] National Laboratory of Solid State Microstructures, Department of Physics, Nanjing University and Collaborative Innovation Center of Advanced Microstructures, Nanjing 210093, People's Republic of China

[4] Departamento de Física, Universidade Federal de Viçosa, 36570-900 Viçosa, Minas Gerais, Brazil

a) Author to whom correspondence should be addressed: antonio.azevedo@ufpe.br

The spin-to-charge conversion phenomena is investigated in a trilayer structure consisting of Co(12 nm)/Sb($t$)/Py(12 nm), where the thickness $t$ of the antimony layer is varied. Using the spin-pumping technique, a pure spin current is injected from both FM layers into the middle layer, the DC voltage is then measured. We observe a spin-to-charge mechanism in the Sb layer that exhibits striking similarities to the inverse Rashba-Edelstein effect (IREE), driven by surface states.

Spin-to-charge conversion is crucial for advancing spintronics, revolutionizing modern electronics. This phenomenon enables the interconversion of spin currents into charge currents and vice versa, enabling highly efficient data processing and storage. Two main mechanisms drive this conversion: the inverse spin Hall effect (ISHE)[1,2] and the inverse Rashba-Edelstein effect (IREE)[3,4]. ISHE arises from scattering caused by spin orbit coupling (SOC) within the bulk of a material, while IREE is an interface effect caused by significant interfacial SOC and an electric field perpendicular to the surface, induced by the breaking of the inversion symmetry. By injecting a pure spin current $\vec{J}_S$ from a ferromagnetic layer (FM) into an adjacent layer, the underlying mechanism can be identified. In bulk materials, the induced charge current density $\vec{J}_C$ due to ISHE is described by,

$$\vec{J}_C = \theta_{SH} \frac{2e}{\hbar} (\vec{J}_S \times \hat{\sigma}),$$ (1)

where, $\vec{J}_C$ is perpendicular to both the spin current density $\vec{J}_S$ and the spin polarization $\hat{\sigma}$. The spin Hall angle $\theta_{SH}$, reflects the efficiency of the conversion process[5,6]. For IREE, the surface charge current density is given by,

$$\vec{J}_C = \frac{e\alpha_R}{\hbar} (\hat{z} \times \vec{S}),$$ (2)

with $\alpha_R$ as the Rashba parameter, $\vec{S}$ representing the non-equilibrium spin density caused by spin injection, and $\hat{z}$ represents the unit vector normal to the interface that aligns with the inversion symmetry-breaking electric field[7,8,9]. During the spin pumping (SP) process, the spin accumulation pointing to $\hat{y}$ direction, leads to the shift of the outer/inner Fermi circles towards the positive/negative $k_x$ direction. Consequently, this generates a net charge current along the positive $k_x$ direction, see Fig. 1 (a)[10]. The charge current generated through IREE is independent of the direction of spin current (Eq. 2), allowing the differentiation between bulk and interfacial conversion by altering the direction of the spin current, as illustrated in Fig. 1 (b). Here, we study the spin-to-charge conversion in Sb thin films, where the spin current can be injected into the middle layer of Sb either from the top via Py (Permalloy = $Ni_{81}Fe_{19}$) or from the bottom via Co. If bulk-driven, the SP signals generated at the middle layer will present inverted signs. If IREE-driven, the signals will remain unchanged, resulting in both signals having the same sign.[8,9]

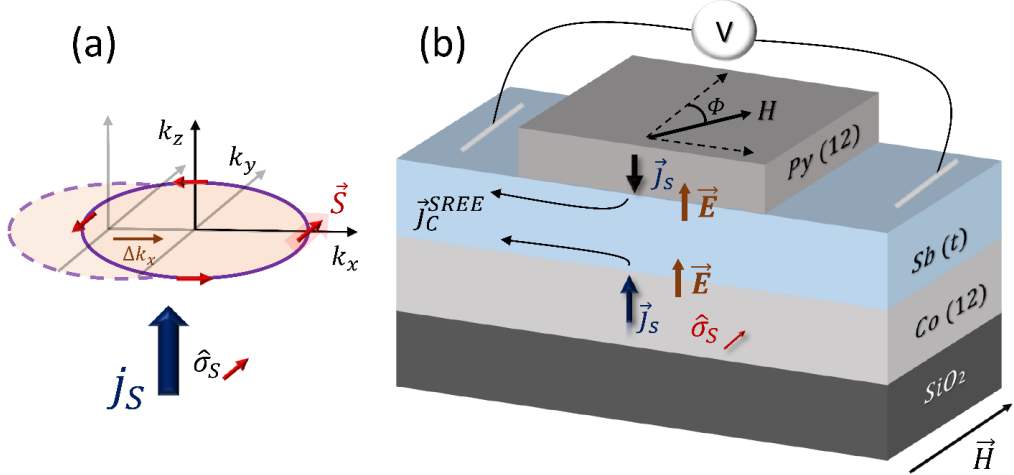

**FIG. 1.** (a) Illustration depicting spin-to-charge conversion process by IREE. (b) Representation of the sample configuration showing the injection of spin currents in both upward and downward directions at the Sb surfaces by Co and Py, respectively, under ferromagnetic resonance conditions.

Using the SP technique driven by ferromagnetic resonance (FMR), we drive spin current from both ferromagnetic layers into the Sb film. This can be achieved through spin pumping since Py and Co have very different resonance fields upon the same microwave excitation. The spin current density pumped by the FM layers in the FMR condition is given by $\vec{J_S} = \frac{\hbar g_{eff}^{\uparrow\downarrow}}{4\pi M^2}\left(\vec{M} \times \frac{\partial \vec{M}}{\partial t}\right)$, where the $g_{eff}^{\uparrow\downarrow}$ is called the spin mixing conductance, $\hbar$ is the reduced Plank's constant and M is the precessing magnetization[11,12].

We deposited samples on Si wafers coated with a 300 nm $SiO_2$ layer via DC sputtering. The substrates were cut into 1.5×4 $mm^2$ pieces, and a 12 nm Co film was sputtered onto them. Then, an antimony film was deposited, followed by sputtering Py in a 1.5×2 $mm^2$ area on top of the Co/Sb stack. A Py island at the center of the upper Sb layer allowed precise electrodes attachments for accurate SP measurements from both the Co and Py layers. FMR condition was achieved using a $TE_{102}$ rectangular microwave cavity, resonating at 9.42 GHz, using an RF power of 110 mW. The sample, mounted on the tip of a PVC rod, is placed at the end of the resonance cavity in a position of maximum rf magnetic field.

For a clear comparison, we performed FMR and SP measurements on a reference sample consisting of a Co/Pt/Py trilayer. Pt is known to exhibit strong SOC, and the primary mechanism responsible for converting spin-to-charge is the ISHE. Co and Py layers were kept constant at 12 nm thickness. The FMR absorption and the SP signal for the Co(12)/Pt(10)/Py(12) sample are shown in Figs. 2(a) and 2(b). The in-plane angle was determined as the angle between the external magnetic field and the axis perpendicular to the line joining the electrodes (Fig. 1(b)).

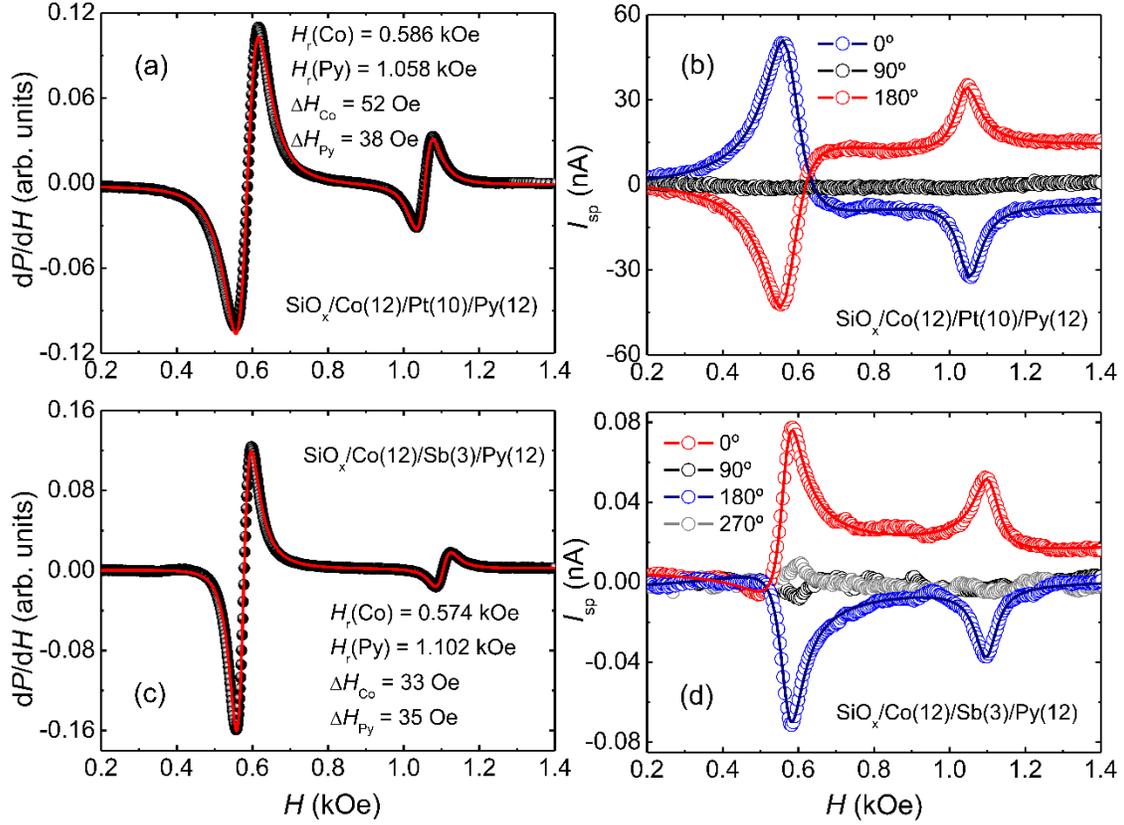

**FIG. 2.** (a) The FMR absorption for the reference sample Co(12)/Pt(10)/Py(12). (b) Spin-pumping result for three different in-plane angles. (c) The FMR absorption of the sample Co(12)/Sb(3)/Py(12). (d) Spin-pumping measurements for four different in-plane angles. Notably, a distinct discrepancy can be observed between Figures (b) and (d) regarding the behavior of the SP signal. In the case of figure (b), the spin-pumping signal follows the trend described by Eq. (1), whereas in the case of figure (d), it adheres to Eq. (2).

In Fig. 2b, the blue symbols show a positive SP signal at a field value of 586 Oe (Co FMR field), while at 1.058 kOe (Py FMR field) the Py SP signal is negative. At reversed field, (red symbols), both SP signals change their sign. This occurs because the Co (Py) layer injects a spin current from bottom to top (top to bottom) but the spin polarization is fixed by the external field direction. This ISHE behavior is governed by Eq. 1. At 90 degrees, the ISHE signal is zero (black symbols). Figs. 2(c) and 2(d) present FMR and SP signals for the Co(12)/Sb(3)/Py(12) sample (the numbers are the layer thicknesses in nm). At Co and Py resonance fields, SP signals (Fig. 2 (d)) exhibit the same sign, according to the Eq. 2. This compelling evidence indicates that the spin-to-charge current conversion is happening through the surface states at both the Co/Sb and Sb/Py interfaces.

The same behavior is observed for different thicknesses of the Sb layer ($t_{Sb}$ = 5, 7, and 15 nm) in Fig. 3. The intensity of the SP signals remains nearly constant regardless of the thickness of the Sb layer, as shown in Fig. 4. This suggests that the surface states present in Sb persist even at larger thickness, like 30 nm (not shown). This implies that there is no significant spin-to-charge conversion due to ISHE present in our Sb films. Also, notice that even though the electrodes were placed in the top surface of the Sb layer, we were able to measure the DC signal originating from the bottom Co/Sb interface. Further confirmation through techniques like angle-resolving photoemission spectroscopy (ARPES)[13] would be necessary to verify if the DC sputtered Sb films are indeed topological insulators.

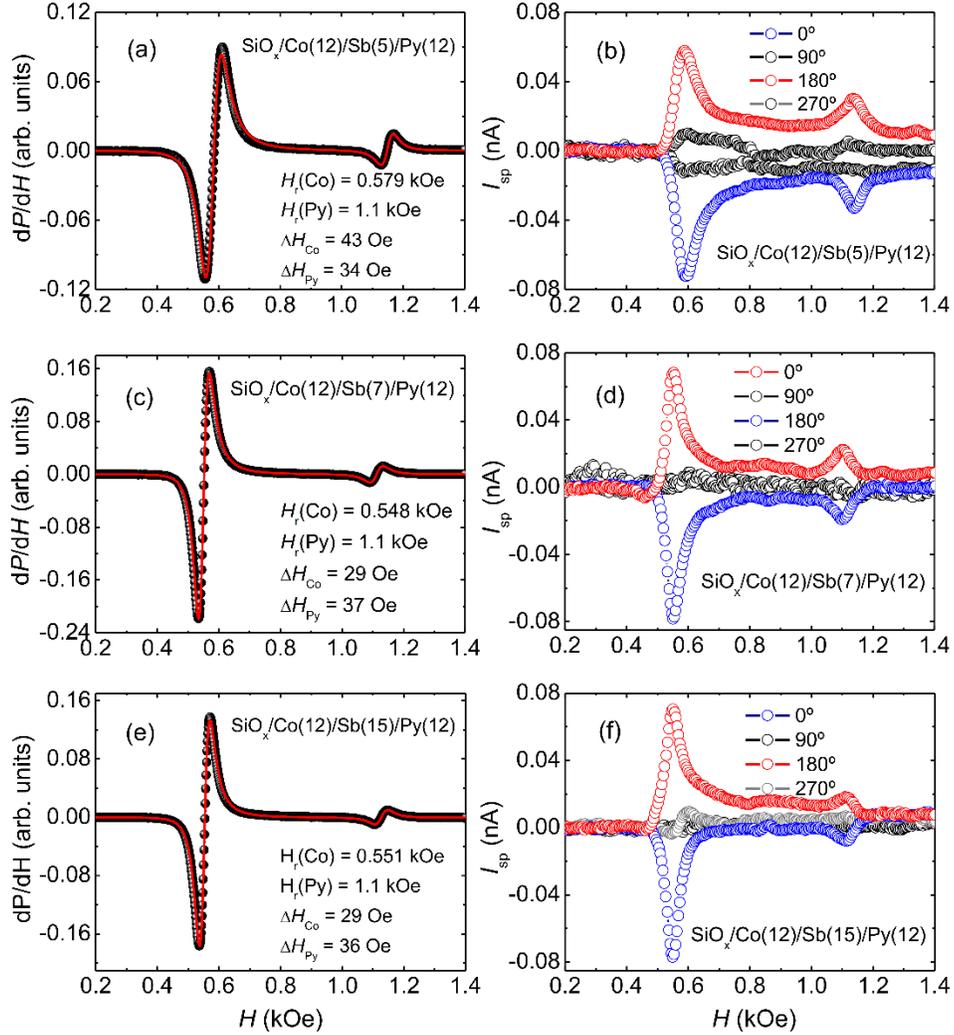

**FIG. 3.** (a, c, e) FMR absorption and (b, d, f) SP signals for samples with varying Sb thickness of 5 nm, 7 nm, and 15 nm. Remarkably, all results exhibit a consistent trend like that observed in Figs. 2(c) and 2(d).

Our study focuses on two interfaces, Co/Sb and Sb/Py, shedding light on the direction of the inversion symmetry-breaking electric field $\left( \vec{E} \parallel \hat{z} \right)$, perpendicular to both interfaces[14]. Our results reveal that the unit vector $\hat{z}$ in equation 1 must point in the same direction at both interfaces. Based on materials working functions, we explain the upward direction of $\vec{E}$ from Co to Sb and from Sb to Py. Referring to the literature, we used the following values for the working functions: $\phi_{Co} \sim 4.1 \ eV$[15], $\phi_{Sb} \sim 4.4 \ eV$[16] and $\phi_{Py} \sim 4.83 \ eV$[17]. By joining the three metals (as shown in Fig. 1(b)) and as the Fermi level is the same for the trilayer sample, electrons migrate from Co to Sb and from Sb to Py. As a result, the electric potential gradients account for the established electric field directions at both interfaces, depicted by the brown arrows in Fig. 1(b).

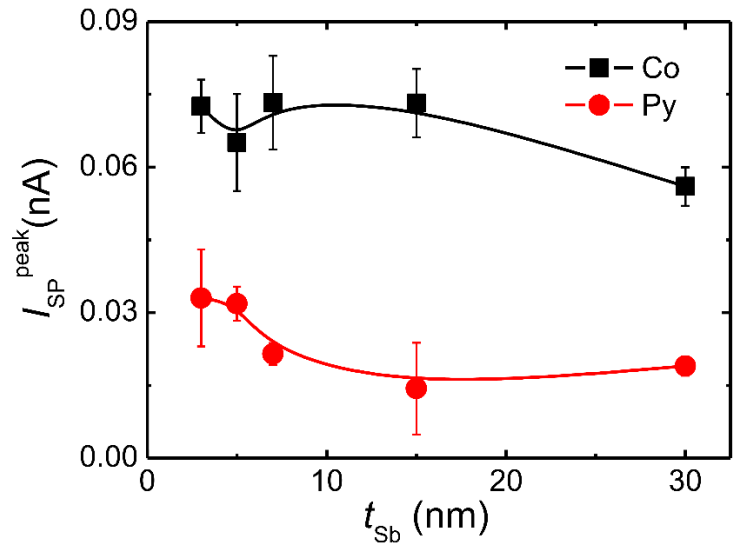

**FIG. 4**: Variation of the $I_{SP}$ peak with the Sb layer thickness. The black (red) symbols represent the current due to the spin injection from the Co (Py) layers. The data points were obtained by averaging the magnitudes of the signals measured at $\phi = 0°$ and $180°$, with error bars indicating the differences. The solid lines are to guide the eyes.

In conclusion, our study reveals that the spin-to-charge conversion in DC sputtered Sb films is primarily attributed to the inverse Rashba-Edelstein-like effect, and the Rashba field remains significant even for Sb thickness as large as 30 nm. Notably, we successfully distinguished the signals originating from IREE and the ISHE by comparing the spin-to-charge conversion processes in Pt and Sb. Using our proposed methodology, we were able to determine the direction of the symmetry-breaking electric field, which aligns parallel to the unit vector $\hat{z}$ as described in Eq. 2. However, further exploration is necessary to understand the underlying physical nature of the surface states discovered in Sb and ascertain whether they exhibit other properties topological insulators[18,19].

### ACKNOWLEDGMENTS

This research is supported by Conselho Nacional de Desenvolvimento Científico e Tecnológico (CNPq), Coordenação de Aperfeiçoamento de Pessoal de Nível Superior (CAPES) (Grant No. 0041/2022), Financiadora de Estudos e Projetos (FINEP), Fundação de Amparo à Ciência e Tecnologia do Estado de Pernambuco (FACEPE), Universidade Federal de Pernambuco, Multiuser Laboratory Facilities of DF-UFPE, Fundação de Amparo à Pesquisa do Estado de Minas Gerais (FAPEMIG) - Rede de Pesquisa em Materiais 2D and Rede de Nanomagnetismo, and INCT of Spintronics and Advanced Magnetic Nanostructures (INCT-SpinNanoMag), CNPq 406836/2022-1.

### DATA AVAILABILITY STATEMENT

The data that support the findings of this study are available from the corresponding author upon reasonable request.